\documentclass[twocolumn,pra, showpacs,amsmath,floatfix]{revtex4}
\usepackage{bm}                      
\usepackage{dcolumn}                 

\newcommand{\AJ  }[3]{Astrophys.~J.                {\bf#1}, #2 (#3).}

\newcommand{\JPB }[3]{J.~Phys. B                   {\bf#1}, #2 (#3).}

\newcommand{\PRA }[3]{Phys.~Rev.~A                 {\bf#1}, #2 (#3).}

\newcommand{\oh}{{\em o}-H$_2$}
\newcommand{\ph}{{\em p}-H$_2$}
\newcommand{\iyr}{{\mathrm{year}}^{-1}}

\newcolumntype{w}[1]{D{.}{.}{#1}}

\begin{document}

\title{Ortho-para transition in molecular hydrogen}

\author{Krzysztof Pachucki}
\email{krp@fuw.edu.pl}
\affiliation{Institute of Theoretical Physics,
             University of Warsaw, Ho{\.z}a 69, 00-681~Warsaw, Poland }

\author{Jacek Komasa}
\email{komasa@man.poznan.pl}
\affiliation{Faculty of Chemistry, 
        A.~Mickiewicz University, Grunwaldzka 6, 60-780~Pozna\'n, Poland }

\date{\today}

\begin{abstract}
The radiative ortho-para transition in the molecular hydrogen is studied.
This highly forbidden transition is very sensitive to 
relativistic and subtle nonadiabatic effects. Our result for the transition 
rate in the ground vibrational level  
$\Gamma(J=1\rightarrow J=0) = 6.20(62)\cdot 10^{-14}\ \iyr$ is
significantly lower in comparison to all the previous approximate
calculations. Experimental detection of such a weak line 
by observation of, for example, the cold interstellar molecular hydrogen is at present unlikely.
\end{abstract}

\maketitle

The hydrogen molecule in the ground electronic state
can exist in a nuclear triplet state  
($S=1$, {\em ortho}-H$_2$) with the odd angular momentum $L$, 
or in a singlet state ($S=0$, {\em para}-H$_2$) with the even $L$. 
The question we  rise is, what are the physical mechanisms
for possible transitions between these two classes of states.
The nonradiative transition, for example, in the interstellar molecular
hydrogen is mostly induced by collisions with atomic H. The corresponding
rates were obtained by Sun and Dalgarno in \cite{dalgarno}. The radiative transition
which is much weaker can in principle take place at sufficiently
low densities and temperatures.
The relativistic spin-orbit interaction (nuclear spin and the electron momenta)
is the most obvious source of this transition as it mixes slightly the 
\oh\ and \ph\ states, see Eq.~(\ref{01}).
This effect has been considered in the original work of 
Raich and Good in \cite{RG64}, although not in a complete and systematic way. 
It has happened that a tiny nonadiabatic correction to the total H$_2$ wave
function significantly changes the theoretical predictions for this rate. 
Moreover, the spin-orbit mixing is not the only one  effect, 
which makes this transition possible.
There are also relativistic corrections
to the $E1$ coupling to the electromagnetic field which is barely known
in literature. These corrections in the context of H$_2$ molecule have
been derived for the first time by Dodelson in \cite{Dod86} using 
the Feinberg-Sucher formalism. Here we rederive this result in a 
much simpler way. Because of the summation over the infinite H$_2$ spectrum,
the calculations of ortho-para transition amplitude are not completely trivial. 
The most elaborate so far calculations by
Raich and Good \cite{RG64} including Dodelson corrections \cite{Dod86}
gave the rate of $1.85(46)\cdot10^{-13}$ year$^{-1}$, 
which did not include all important contributions.
The purpose of this work is to present a complete theoretical description
of the radiative ortho-para conversion of H$_2$ molecule, including the results 
of direct numerical calculations of the transition probability 
for the lowest rotational levels.

The interaction of an arbitrary molecule with the electromagnetic field,
whose characteristic wavelength is much larger than the size of this molecule,
including the relevant spin-orbit interaction is \cite{Pac07}
\begin{eqnarray}
\delta H &=& 
-\sum_A\,e_A\,\vec x_A\cdot\vec E 
-\sum_b\,e_b\,\vec x_b\cdot\vec E 
-\sum_A\frac{e_A}{2\,m_A}
\nonumber \\ &&\times\biggl[
g_A\,x_A^i\,s_A^j\,B^j_{,i} + (g_A-1)\,\vec
s_A\times\vec x_A\cdot\partial_t\vec E\biggr]
\nonumber \\ &&
+\sum_{A,b}\frac{e_A\,e_b}{4\,\pi}\,
\frac{1}{2\,x^3_{Ab}}\,\biggl[\frac{g_A}{m_A\,m_b}\,
\vec s_A\cdot\vec x_{Ab}\times\vec p_b
\nonumber \\ &&
-\frac{(g_A-1)}{m_A^2}\,
\vec s_A\cdot\vec x_{Ab}\times\vec p_A \biggr].\label{01}
\end{eqnarray}
In the above, $A, b$ are the indices of the nuclei and electrons respectively,
$x_A, x_b$ are the coordinates of the nuclei and electrons with respect to 
the mass center, $\vec x_{Ab} = \vec x_A - \vec x_b$, and $g_A$
is the  nuclear g-factor. Moreover, the electromagnetic fields 
$\vec E, \vec B$ and their derivatives 
are assumed in Eq.~(\ref{01}) to be at the mass center.

For \oh\ (the first excited rotational state) the nuclear spin $S=1$
couples to the orbital angular momentum $J=1$ of the nuclei, giving the
total angular momentum characterized by quantum numbers $F=0,1,2$.  In the
\ph\ (the ground rotational state) the total angular momentum is $F=0$,
therefore the one photon ortho-para transition from any other $F=0$ level
is strictly forbidden, while $F=1$
level decays by $E1$ transition and $F=2$ level decays by the $M2$ transition.

Let us first consider the $M2$ transition from $F=2$ level of \oh, 
to $F=0$ level of \ph. This transition comes from the following interaction with the
electromagnetic field, which is obtained from  Eq.~(\ref{01})
\begin{equation}
\delta H = 
-\frac{e\,g_p}{4\,m_p}\,(s_A^j - s_B^j)\,{\cal R}^i\,B^j_{,i}\,, \label{02}
\end{equation}
with ${\cal R} = x_A-x_B$ and the proton g-factor $g_p=5.585\,694\,713(46)$ \cite{nist}.
From this Hamiltonian one obtains the transition rate $\Gamma(M2)\equiv\Gamma_2$
\begin{eqnarray}
\Gamma_2 &=& 2\,\alpha\,\omega\;
\frac{1}{5}\sum_{m_F}\biggl|\biggl\langle {}^1\Sigma_0\biggl| 
\frac{g_p}{4\,m_p}\,(\vec k\cdot\vec {\cal R})\,
(\vec s_A - \vec s_B)\times\vec k
\nonumber \\ &&
\biggr|{}^3\Pi_2,m_F\biggr\rangle\biggr|^2
\nonumber \\ &\approx&
\frac{1}{120}\,\alpha\,\omega^5\,\biggl(\frac{g_p\,{\cal
    R}_0}{m_p}\biggr)^2 \,. \label{03}
\end{eqnarray}
where ${\cal R}_0$ is the average distance between protons, and the last
equation holds for the nuclear H$_2$ wave function which is strongly 
peaked around ${\cal R}_0$, as it is the case for the lowest 
rovibrational levels.

The calculation of the rate for $E1$ transition from $F=1$ level of \oh\ to
$F=0$ level of \ph\ is more complicated as it includes 
corrections to the wave function coming  from the spin-orbit interaction. 
Since it is $\Delta F=1$ transition, the 
operators in the interaction Hamiltonian in Eq.~(\ref{01}) can be
simplified, namely $x_A^i\,s_A^j \rightarrow \epsilon^{ijk}\,(\vec
x_A\times\vec s_A)^k/2$ and this Hamiltonian becomes
\begin{eqnarray}
\delta H &=& H_{LS} + 
e\,(\vec x_{1C}+\vec x_{2C}) \cdot\vec E 
\nonumber \\ &&
+\frac{e}{4\,m_p}\,\biggl(\frac{g_p}{2}-1\biggr)\,\vec{\cal R}\times
(\vec s_A - \vec s_B)\cdot\partial_t\vec E\,,\label{04}\\
H_{LS} &=& -\frac{i}{2}\,\vec H_{LS}\cdot(\vec s_A-\vec s_B), \\
\vec H_{LS} &=& \frac{g_p\,\alpha}{2\,m_p\,m}\,\vec h_1-
\frac{(g_p-1)\,\alpha}{2\,m_p^2}\,\vec h_2\times\vec\nabla_{\cal R}\\
\vec h_1 &=&\biggl(\frac{\vec x_{1A}}{x^3_{1A}} - \frac{\vec x_{1B}}{x^3_{1B}}
\biggr)\times\vec \nabla_1 +
\biggl(\frac{\vec x_{2A}}{x^3_{2A}} - \frac{\vec x_{2B}}{x^3_{2B}}
\biggr)\times\vec \nabla_2,\nonumber \\ \\
\vec h_2 &=& 
\frac{\vec x_{1A}}{x_{1A}^3}+\frac{\vec x_{1B}}{x_{1B}^3}+
\frac{\vec x_{2A}}{x_{2A}^3}+\frac{\vec x_{2B}}{x_{2B}^3}\,.
\end{eqnarray}
The resulting transition rate $\Gamma(E1)\equiv\Gamma_1$ is
\begin{eqnarray}
\Gamma_1 &=& \frac{4}{3}\,\alpha\,\omega\;\frac{1}{3}\,\sum_{m_F}
 |\langle ^1\Sigma_0|\vec Q\times\frac{(\vec s_A-\vec
  s_B)}{2}|^3\Pi_1,m_F\rangle|^2 \nonumber \\ &=&
\frac{2}{9}\,\alpha\,\omega|\langle\Sigma|Q^k|\Pi^k\rangle|^2\,,
\label{05}
\end{eqnarray}
where
\begin{eqnarray}
Q^k &=& \frac{1}{2\,m_p}\,\biggl(\frac{g_p}{2}-1\biggr)\,\omega^2\,{\cal R}^k
\nonumber \\ &&
-\frac{\omega}{2}\,\epsilon^{ikj}\,(x_{1C}^i+x_{2C}^i)\,
\frac{1}{E_{\Pi}-H}\,H_{LS}^j \nonumber \\ &&
-\frac{\omega}{2}\,\epsilon^{ikj}\,H_{LS}^j\,
\frac{1}{E_{\Sigma}-H}\,(x_{1C}^i+x_{2C}^i),\label{06}
\end{eqnarray}
and $H$ is the 4-body nonrelativistic Hamiltonian. 
In order to simplify the evaluation of $Q^k$ we perform the adiabatic expansion,
namely the expansion of the resolvent in the kinetic energy of the nuclei
\begin{eqnarray}
\frac{1}{E_{\Pi}-H} &=& \frac{1}{E_{\rm BO}-H_{\rm BO}}
-\frac{1}{E_{\rm BO}-H_{\rm BO}}\,(\delta E_{\Pi}-\delta H_M)
\nonumber \\ &&
\times\frac{1}{E_{\rm BO}-H_{\rm BO}}+\ldots\label{07}
\end{eqnarray}
similarly for $\frac{1}{E_{\Sigma}-H}$,
and the expansion of the wave function
\begin{eqnarray}
\phi_{\Sigma} &=& \psi(\vec x_{1C},\vec x_{2C};{\cal R})\,
\lambda_0({\cal R})/\sqrt{4\,\pi} + \delta\phi_\Sigma, \label{08}\\
\phi_{\Pi}^k &=& 
\psi(\vec x_{1C},\vec x_{2C};{\cal R})\,\lambda_1^k(\vec{\cal R})/\sqrt{4\,\pi}
 + \delta\phi_\Pi, \label{09}
\end{eqnarray} 
with $\lambda_1^k = \lambda_1\,{\cal R}^k/{\cal R}$ and with normalization
\begin{equation}
\int d{\cal R}\,{\cal R}^2\,\lambda_0^{2}({\cal R}) = \int
d{\cal R}\,{\cal R}^2\,\lambda_1^{2}({\cal R})=1\,.\label{10}
\end{equation}
While the exact nonadiabatic correction is unknown, we need only the first order 
$m/m_p$ part of the correction, which explicitly depends on 
the nuclear state \cite{krp}
\begin{eqnarray}
\delta\phi_\Sigma &=& -\frac{2}{m_p}\,\frac{1}{E_{BO}-H_{BO}}\,
\nabla^l_{\cal R}\psi\,\nabla^l\lambda_0/\sqrt{4\,\pi}, \label{11}\\
\delta\phi_\Pi^k  &=& -\frac{2}{m_p}\,\frac{1}{E_{BO}-H_{BO}}\,
\nabla^l_{\cal R}\psi\,\nabla^l\lambda_1^k/\sqrt{4\,\pi}.\label{12}
\end{eqnarray} 

We introduce now the perturbed electronic wave functions
\begin{subequations}\label{psi123}
\begin{eqnarray}
\psi_1^i &=& \frac{1}{E_{\rm BO}-H_{\rm BO}}\,(x_{1C}^i+x_{2C}^i)\,\psi,
\label{psi1}\\
\psi_2^j &=& \frac{1}{E_{\rm BO}-H_{\rm BO}}\,h_1^j\,
\psi,\label{psi2}\\
\psi_3^l &=&  \frac{1}{E_{\rm BO}-H_{\rm BO}}\,\nabla^l_{\cal R}\,\psi,
\label{psi3}
\end{eqnarray}
\end{subequations}
to simplify  the matrix elements of  $Q^k$ in Eq.~(\ref{06}) 
\begin{eqnarray}
\langle \Sigma|Q^k|\Pi^k\rangle &=&
\frac{1}{2\,m_p}\,\biggl(\frac{g_p}{2}-1\biggr)\,\omega^2\,{\cal R}_0
-\frac{\omega\,(g_p-1)}{m_p^2\,{\cal R}_0}
\label{16}\\ &+&
\omega^2\,\frac{g_p\,\alpha}{4\,m_p\,m}\,X_1 
-\omega\,\frac{g_p\,\alpha}{4\,m_p^2\,m\,{\cal R}_0}\,(X_2+2\,X_3)
\nonumber 
\end{eqnarray}
where we used the commutator
\begin{equation}
i\,\bigl[p_1^k+p_2^k,H_{BO}-E_{BO}\bigr] = \alpha\,h_2^k,
\end{equation}
and
\begin{subequations}\label{EXid}
\begin{eqnarray}
X_1 &=& \epsilon^{ikj}\,n^k\,\bigl\langle\psi_1^i\,|\,
\psi_2^j\bigr\rangle_{{\cal R}_0}, \\
X_2 &=& (\delta^{kl}-n^k\,n^l)\,\epsilon^{ikj}\nonumber \\ &&
\times\Bigl[\bigl\langle\partial^l_{\cal R}\psi_1^i\,|\,\psi_2^j\bigr\rangle-
\bigl\langle\psi_1^i\,|\,
\partial^l_{\cal R}\psi_2^j\bigr\rangle\Bigr]_{{\cal R}_0},\\
X_3 &=& (\delta^{kl}-n^k\,n^l)\,\epsilon^{ikj}\,
\Bigl[\langle\psi_3^l|(x_{1C}^i+x_{2C}^i)|\psi_2^j\rangle
\nonumber \\ &&
+\langle\psi_3^l|h_1^j|\psi_1^i\rangle\Bigr]_{{\cal R}_0},
\end{eqnarray}
\end{subequations}
with $\vec n \equiv \vec{\cal R}_0/{\cal R}_0$. 
One notes that derivatives of nonlinear and linear parameters in 
$\psi$, see Eq.~(\ref{GG}), with respect to 
$\cal R$ do not contribute to the above matrix elements, 
which significantly simplifies the numerical computations. 

Results for $X_i$ can be expressed in terms of dimensionless factors $F_i$
\begin{subequations}\label{EXif}
\begin{eqnarray}
X_1 &=& -\frac{9}{m^2\,\alpha^4\,{\cal R}^2}\,F_1(m\,\alpha\,{\cal R}), \\
X_2 &=& \frac{9}{m^2\,\alpha^4\,{\cal R}^3}\,F_2(m\,\alpha\,{\cal R}),  \\
X_3 &=& -\frac{2\,m}{\alpha}\,F_3(m\,\alpha\,{\cal R}),
\end{eqnarray}
\end{subequations}
which are chosen in such a way, that 
$F_i(m\,\alpha\,{\cal R})$ vanish
at ${\cal R}=0$ and approach 1 for ${\cal R}\to\infty$.
However, in the case of  $F_2$ this large ${\cal R}$ limit 
is only a rough approximation, since we have not been able to perform this
limit analytically.
We can now return to Eq.~(\ref{16}) and obtain a compact formula
for the matrix element in the transition rate $\Gamma_1$ of Eq.~(\ref{05})
\begin{eqnarray}\label{Qkf}
\langle \Sigma|Q^k|\Pi^k\rangle &=& \frac{\omega^2\,{\cal R}_0}{2\,m_p}\,
\biggl[\biggl(\frac{g_p}{2}-1\biggr)
-\frac{9}{2}\,\frac{g_p\,F_1}{(m\,\alpha\,{\cal R}_0)^3}\biggr]
 \\ &-&
\frac{\omega}{m_p^2\,{\cal R}_0}\,\biggl[(g_p-1)-g_p\,F_3 +\frac{9}{4}\,
\frac{g_p\,F_2}{(m\,\alpha\,{\cal R}_0)^3}\biggr].\nonumber
\end{eqnarray}

Numerical evaluation of the $\Gamma_2$ rate according to formula
(\ref{03}) is straightforward. To obtain the ortho-para energy spacing 
$\omega$ and the average internuclear distance ${\cal R}_0$, 
we employed the accurate Ko{\l}os-LeRoy-Schwartz interaction
potential \cite{KLS87} which includes the adiabatic and relativistic energy
corrections. With this potential we solved numerically the
radial Schr{\"o}dinger equation to obtain the energies and wave functions
corresponding to the lowest ortho and para levels. The
numerical values used here are $\omega=2\pi\cdot 118.49$~cm$^{-1}$ and
${\cal R}_0 =1.449$ au, and the resulting $M2$ transition rate 
with physical constants from Ref. \cite{nist} is
\begin{equation}
\Gamma_2=1.07(1)\cdot 10^{-14}\,\iyr\,.\label{20}
\end{equation}

The accurate evaluation of $\Gamma_1$ and the corresponding
electronic matrix elements $F_i$ in Eqs.~(\ref{EXif}) is a challenging task.   
We have represented the electronic ground state wave function as well as 
the first order perturbed functions defined by Eqs.~(\ref{psi123}),
in the form of properly symmetrized linear combinations 
$\psi=\sum_k c_k \hat{P}_\mathrm{g,u} \phi_k$
of Gaussian geminals
\begin{eqnarray}
\lefteqn{\phi_k=\Xi_{k}} \label{GG} \\ &&
\times\exp\left(-\alpha_k\, x_{1A}^2-\beta_k\, x_{1B}^2-\zeta_k\, x_{2A}^2
	-\eta_k\, x_{2B}^2 -\gamma_k\, x_{12}^2\right).\nonumber
\end{eqnarray}
The projection operators  
\begin{equation}\label{Pgu}
\hat{P}_\mathrm{g,u} = \frac{1}{4}\,(1+\hat P_{12})\,(1\pm\hat{\imath})
\end{equation}
ensure the proper symmetry with respect to the exchange of the electrons and
with respect to the inversion operation, yielding singlet {\em gerade} or 
{\em ungerade} functions. Required $\Sigma^{^+}$, $\Sigma^{^-}$, or $\Pi$
symmetry of the electronic wave function
was imposed by the Cartesian prefactor $\Xi_{k}$. The linear and 
the nonlinear parameters were optimized variationally with the goal 
function being the ground state energy in the case of the unperturbed 
wave function $\psi$ or pertinent Hylleraas functional
\begin{equation}\label{J}
{\cal J}[\psi_k^i]=\langle\psi_k^i|E_{\rm BO}-H_{\rm BO}| \psi_k^i\rangle+2\langle\psi|
\hat{\cal O}|\psi_k^i\rangle
\end{equation}
in the case of the perturbed functions $\psi_k^i$.
Table~\ref{T1} shows explicitly the elements defining particular functions with
the assumption that the molecule is placed along the Cartesian $X$ axis.
\begin{table}[ht]\renewcommand{\arraystretch}{1.2}
\caption{\label{T1} The definitions of the functions used in the computations.}
\begin{ruledtabular}
\begin{tabular*}{0.47\textwidth}{c@{\extracolsep{\fill}}ccc}
&$\Xi_{k}$  & $\hat{P}$, Eq.~(\ref{Pgu}) & $\hat{\cal O}$, Eq.~(\ref{J})\\
\hline
$\psi$     & 1               &   gerade & $-$       \\
$\psi_1^x$ & $x_1$, $x_2$    & ungerade & $x_1+x_2$ \\
$\psi_1^y$ & $y_1$, $y_2$    & ungerade & $y_1+y_2$ \\
$\psi_1^z$ & $z_1$, $z_2$    & ungerade & $z_1+z_2$ \\
$\psi_2^x$ & $y_1z_2-y_2z_1$ & ungerade & $h_1^x$   \\
$\psi_2^z$ & $y_1$, $y_2$    & ungerade & $h_1^z$   \\
$\psi_3^y$ & $y_1$, $y_2$    &   gerade & $\partial^y_{\cal R}$ \\
\end{tabular*}
\end{ruledtabular}
\end{table}
The unperturbed wave function has been expanded in 600-term basis set 
which enables the electronic ground state energy to be obtained with an error of only
$3\cdot 10^{-9}$ a.u. A 1200-term expansions have been employed to represent 
the perturbed functions. Values of the ${\cal J}$ functionals corresponding 
to the optimum parameters are displayed in Table~\ref{T2}.

The general formulas (\ref{EXid}), in the particular case of the molecule
oriented along the $X$ axis, can be explicitly written as follows
\begin{subequations}\label{EXie}
\begin{eqnarray}
X_1 &=& -2\bigl\langle\psi_1^y\,|\,\psi_2^z\bigr\rangle_{{\cal R}_0}, \\
X_2 &=& 
4\Bigl[\bigl\langle\partial^y_{\cal R}\psi_1^x\,|\,\psi_2^z\bigr\rangle
-\bigl\langle\partial^y_{\cal R}\psi_1^z\,|\,
\psi_2^x\bigr\rangle\Bigr]_{{\cal R}_0},\\
X_3 &=& 
2\Bigl[\langle\psi_3^y|(x_{1C}^x+x_{2C}^x)|\psi_2^z\rangle 
-\langle\psi_3^y|(x_{1C}^z+x_{2C}^z)|\psi_2^x\rangle  \nonumber\\ &&
+\langle\psi_3^y|h_1^z|\psi_1^x\rangle-\langle\psi_3^y|h_1^x|\psi_1^z\rangle
\Bigr]_{{\cal R}_0}.
\end{eqnarray}
\end{subequations}
Table~\ref{T2} contains all the expectation values
appearing in Eqs.~(\ref{EXie}) as well as the final $X_i$ and $F_i$ values computed at ${\cal R}_0=1.449$ bohr. To check the correctness of our codes we
performed additional calculations at large internuclear distance (${\cal
  R}=12.0$ bohr) and compared the resulting goal functions and the expectation values with analytically derived asymptotic values. This comparison is presented in Table~\ref{T2}.
\begin{table}[htb]\renewcommand{\arraystretch}{1.2}
\caption{\label{T2} Numerical values of the optimum goal functions 
and the expectation values comprising the $X_i$ factors. The asymptotic value
for $F_2$ is approximate.}
\begin{ruledtabular}
\begin{tabular*}{0.45\textwidth}{c@{\extracolsep{\fill}}w{3.10}w{3.10}w{3.4}}
 & \multicolumn{1}{c}{${\cal R}_0=1.449$} & 
\multicolumn{1}{c}{${\cal R}=12.0$} & \multicolumn{1}{c}{Asymp.} \\
\hline
$E_{\mathrm{BO}}$ & -1.174\,073\,569 & -1.000\,002\,546 & -1.0 \\
${\cal J}[\psi_1^x]$ & -3.3582 & -4.5241 & -4.5 \\
${\cal J}[\psi_1^y]$ & -2.3684 & -4.4885 & -4.5 \\
${\cal J}[\psi_2^x]$ & -7.87\times 10^{-3} & -2.22\times 10^{-7} & 0.0 \\
${\cal J}[\psi_2^z]$ & -0.4925 & -5.06\times 10^{-5} &  0.0\\
${\cal J}[\psi_3^y]$ & -2.90\times 10^{-2} & -0.2500 &  -0.25\\[1ex]
$\bigl\langle\psi_1^y\,|\,\psi_2^z\bigr\rangle$ & 
							    0.7824 &  0.0312 &  \\
$X_1$\hspace{\fill} & 	-1.5649 & -0.0623 &  \\
$F_1$\hspace{\fill} & 	 0.3651 &  0.9975 & 1.0 \\[1ex]
$\bigl\langle\partial^y_{\cal R}\psi_1^x\,|\,\psi_2^z\bigr\rangle$ &
								 0.7467 &  0.0014 &   \\
$\bigl\langle\psi_1^z\,|\,\partial^y_{\cal R}\psi_2^x\bigr\rangle$ & 
								 0.0034 &  0.0002 &   \\
$X_2$\hspace{\fill} &    3.0003 &  0.0062 &  \\
$F_2$\hspace{\fill} &    1.0142 &  1.1989 & 1.0 \\[1ex]
$\langle\psi_3^y|(x_{1C}^x\!+\!x_{2C}^x)|\psi_2^z\rangle$ & 
								 0.0707 & -0.0029 & \\
$\langle\psi_3^y|(x_{1C}^z\!+\!x_{2C}^z)|\psi_2^x\rangle$ & 
								-0.0010 & -0.0001 & \\
$\langle\psi_3^y|h_1^z|\psi_1^x\rangle$               & 
								-0.2579 & -0.5026 & \\
$\langle\psi_3^y|h_1^x|\psi_1^z\rangle$               & 
								 0.1066 &  0.4980 & \\
$X_3$\hspace{\fill} &   -0.3984 & -2.0037 & \\
$F_3$\hspace{\fill} &    0.1992 &  1.0019 & 1.0 \\
\end{tabular*}
\end{ruledtabular}
\end{table}

Using Eqs. (\ref{05}), (\ref{Qkf}), and the Table \ref{T2}
one obtains the numerical value for the $E1$ transition rate
\begin{equation}
\Gamma_1=1.68(17)\cdot 10^{-13}\ \iyr\,,\label{23}
\end{equation}
and finally the rate averaged over the total angular momentum $F$
\begin{equation}
\Gamma = (5\,\Gamma_2 + 3\,\Gamma_1)/9 = 6.20(62)\cdot 10^{-14}\ \iyr\,.
\label{24}
\end{equation}

Our result for the averaged transition rate is in disagreement with
the result of Dodelson \cite{Dod86}, $\Gamma = 1.85(46)\cdot 10^{-13}\,\iyr$,
which is in turn based on the former work of Raich and Good \cite{RG64}
and included direct coupling of nuclear spin to the radiation field.
We confirm in this work the existence of these additional couplings,
which here are expressed by the 3rd term in Eq.~(\ref{01}). 
In our opinion, the difference from our result 
is due to the omission of the $M2$ transition, the omission of the 
nonadiabatic contributions corresponding to $X_2$ and $X_3$ in Eq.~(\ref{EXid}),
less accurate $\omega$, and due to a lower accuracy of the numerical calculation
of the matrix elements in Ref.~\cite{RG64}. In particular, without $X_3$
the overall rate $\Gamma$ would be about 24\% larger.

The possibility of the experimental detection of the {\em o-p} H$_2$ line
is questionable. Much stronger $E2$ lines have already been observed
at the {\sl Infrared Space Observatory} ISO and served for estimation of 
the temperature of interstellar hydrogen clouds and of the ratio
of abundance \oh\ to \ph, which sometimes differs significantly from
the equilibrium one \cite{exp1}. The much weaker $E1$ line 
has not been observed yet. In fact there is a potential opportunity 
related with the Herschel Space Observatory to be launched in 2008 \cite{herschel}. 
Its spectral range covers the {\em o-p} line at $84.4\,\mu$m, but 
its resolution is, probably, not high enough at this wavelength.

\section*{Acknowledgments}
K.P. wishes to acknowledge interesting discussions with Krzysztof Meissner,
and thanks the Laboratoire Kastler Brossel in Paris for a kind hospitality
during his stay, when this work was written.

\end{document}